# Multiferroic Iron Oxide Thin Films at Room-Temperature

*Martí Gich*[1], Ignasi Fina[2], Alessio Morelli[2,3], Florencio Sánchez[1], Marin Alexe[4], Jaume Gàzquez[1], Josep Fontcuberta[1] and Anna Roig[1]*

[1]Institut de Ciència de Materials de Barcelona (ICMAB-CSIC), Campus UAB, 08193, Bellaterra, Catalonia, Spain.

[2]Max Planck Institute of Microstructure Physics, Weinberg 2, D-06120, Halle/Saale, Germany.

[3]Department of Dielectrics, Institute of Physics ASCR, 18221, Prague, Czech Republic.

[4]University of Warwick, Department of Physics, CV47AL, Coventry, United Kingdom.

M. Gich and I. Fina contributed equally to this work

*E-mail: mgich@icmab.es

## Abstract

In spite of being highly relevant for the development of a new generation of information storage devices, not many single-phase materials displaying magnetic and ferroelectric orders above room temperature are known. Moreover, these uncommon materials typically display insignificant values of the remanent moment in one of the ferroic orders or are complex multicomponent oxides which will be very challenging to integrate in devices. Here we report on the strategy to stabilize the metastable ε-$Fe_2O_3$ in thin film form, and we show that besides its already known ferrimagnetic nature, the films are also ferroelectric at 300 K with a remanent polarization of 1 μC/$cm^2$. The film polarization shows long retention times and can be switched under small applied voltages. These characteristics make of ε-$Fe_2O_3$ the first single-ion transition-metal oxide which is ferro(ferri)magnetic and ferroelectric at room temperature. The simple composition of this new multiferroic oxide and the discovery of a robust path for its thin film growth may boost the exploitation of ε-$Fe_2O_3$ in novel devices.

**Introduction**

Magnetoelectric multiferroics are currently being intensively studied owing to its rich and fascinating physics and the promise of a novel generation of devices combining the best characteristics of ferromagnetic and ferroelectric materials.[1] In the field of information storage, these materials could bring significant advances towards memories displaying fast and low-power write operation combined with a non-destructive reading or increased computing capacities thanks to the storage of more than one bit per cell.[2] To make these applications a reality one needs materials with sizeable magnetization and electric polarization above room temperature as well as strong coupling between the electric and magnetic dipole orders. At ambient conditions, such combination of properties is extremely unusual in single phase materials and most of the reported examples have the drawback of displaying insignificant values of the remanent moment in one of the ferroic orders. Thus, the application-driven research in multiferroics has focused in alternative materials like composites from magnetic and ferroelectric materials[3] coupled by mechanical strain,[4] electric field effects[5] or exchange bias at the interfaces.[6] Recently, the hopes of achieving practical devices based on single-phase materials have been renewed by the discovery of room-temperature magnetoelectric effects in ferromagnetic multicomponent transition metal oxides such as $Sr_3Co_2Fe_{24}O_{41}$,[7] $Sr_3Co_2Ti_2Fe_8O_{19}$,[8] and in $[Pb(Zr_{0.53}Ti_{0.47})O_3]_{0.6}$-$[Pb(Fe_{0.5}Ta_{0.5})O_3]_{0.4}$ solid solutions.[9] These very complex oxides have only been obtained as polycrystalline powders, even though in reference [9a] the material characterization was entirely performed on a micron-sized single crystal. One can foresee that its fabrication as epitaxial thin films with the control of stoichiometry and crystallinity needed for its integration in practical devices may prove to be extremely challenging. More attractive is the case of GaFeO$_3$, which has also been recently claimed to be ferroelectric at room temperature;[10] however, the magnetic Curie temperature of GaFeO$_3$ is below room temperature[10a, 11] thus precluding room-temperature multiferroicity. Although recent reports[12] have shown that by Fe enrichment (Ga$_{1-x}$Fe$_{1+x}$O$_3$), the Curie temperature could be pushed to around room temperature at expenses of a more intricate stoichiometry, cationic distribution and ferroelectricity stability issues. Thus, room temperature multiferroics composed of a small number of atomic species and having a structure lattice-matching with commonly available single crystalline substrates present a clear advantage. Interestingly, transition metal monoxides such as Fe$_3$O$_4$[13] and CuO[14] are low-temperature multiferroics which point out the necessity of investigating the existence of room temperature multiferroicity among other simple oxides. In this context, we directed our attention towards the polar ε-Fe$_2$O$_3$ (space group Pna2$_1$). ε-Fe$_2$O$_3$ is a robust room-temperature insulating ferrimagnet with a Curie temperature of about 500 K[15] which is driving current interest for possible applications.[16] It is isostructural[17] with GaFeO$_3$, one of the first reported magnetoelectric oxides[18] and multiferroic near room temperature as mentioned.[10] Therefore, due to its high Curie temperature, ε-Fe$_2$O$_3$ appears as an ideal candidate as long as its ferroelectric character could be demonstrated. In fact, our earlier experimental observations provided hints that a magnetoelectric coupling may exist,[19] which were also supported by the presence of a substantial room-temperature spin-orbit coupling[20] similarly to GaFeO$_3$,[21] and electromagnonic excitations.[22] However, ε-Fe$_2$O$_3$ is an elusive metastable phase in the Fe-O phase diagram. Indeed, so far, ε-Fe$_2$O$_3$ is only reproducibly prepared in nanoparticle form.[23] Recently, we reported that the stabilization of ε-Fe$_2$O$_3$ phase in the form of thin epitaxial films could be achieved under stringent conditions.[24] Aiming at growing films of improved quality that would enable a thorough investigation of its dielectric properties and to find a path for reliable epitaxial growth of ε-Fe$_2$O$_3$, we explore here a strategy consisting on using an isostructural epitaxial template layer acting as seed layer. The most simple and natural choice would be the use of an oxide isostructural to ε-Fe$_2$O$_3$ with a stable phase under the typical conditions for epitaxial oxide growth. To this regard, AlFeO$_3$ appears to be ideal, because in addition of being isostructural[25] and entropically stabilized over its κ-Al$_2$O$_3$ and ε-Fe$_2$O$_3$ constituents in the harsh film growth conditions,[26] it presents a reduced mismatch on SrTiO$_3$, the most used single crystalline perovskite substrate. We note that AlFeO$_3$ is ferrimagnetic only below 250 K[25] and thus its magnetic response at room-temperature does not mask that of ferrimagnetic ε-Fe$_2$O$_3$ film.

Here we present fully epitaxial ε-Fe$_2$O$_3$ thin films grown on AlFeO$_3$ buffered Nb-doped SrTiO$_3$(111) single crystalline substrates (STO(111)). We show the room-temperature ferrimagnetism and ferroelectricity of the films by in-depth investigation of the dielectric and magnetic properties, as well as local nano-scale piezoelectric activity and ferroelectric switching using piezoresponse force microscopy (PFM). Our ferrimagnetic ε-Fe$_2$O$_3$ thin films display a switchable polarization with a remanent polarization of about 1 μC/cm$^2$ and a remanent magnetization of about 40 emu/ cm$^3$. A small magnetocapacitance response is also detected, revealing the existence of magnetoelectric coupling. This makes ε-Fe$_2$O$_3$ the only single-ion oxide room-temperature multiferroic that has been experimentally demonstrated so far.

Since ε-Fe$_2$O$_3$ is a metastable phase under normal conditions, we first focus on the growth and structural properties of the films. Epitaxial growth of ε-Fe$_2$O$_3$ has been performed by pulsed laser deposition (PLD) on Nb:STO(111) using AlFeO$_3$ as buffer layer.

**Results and Discussion**

Figure 1 shows the X-ray diffraction (XRD) θ-2θ scans of AlFeO$_3$ (black) and ε-Fe$_2$O$_3$/AlFeO$_3$ (red) films grown on Nb:STO(111) substrates. The X-ray diffraction investigations indicate the existence of a (00$l$) out-of plane texture of the grown films. The AlFeO$_3$ reflections occur at 2θ angles higher than those of ε-Fe$_2$O$_3$ due to the smaller cationic size of Al$^{3+}$ compared to Fe$^{3+}$. Indeed, the cell parameters of these isomorphs are: $a_{AFO}$=4.984 Å, $b_{AFO}$=8.554 Å, $c_{AFO}$=9.241 Å for bulk AlFeO$_3$ powders[25] and $a_\varepsilon$=5.1019 Å, $b_\varepsilon$=8.7807 Å, $c_\varepsilon$=9.4661 Å for ε-Fe$_2$O$_3$ nanoparticles.[27] As a result, the (004) and (008) reflections of AlFeO$_3$ are respectively masked by the substrate (111) and (222) peaks. In contrast, in the ε-Fe$_2$O$_3$/AlFeO$_3$ sample the entire set of ε-Fe$_2$O$_3$ (00$l$) reflections is visible. The out of plane cell parameters evaluated from the (006) reflections of Figure 1(a) are 9.271(5) Å for the AlFeO$_3$ film and 9.450(3) Å for the ε-Fe$_2$O$_3$ film, which are in agreement with the bulk values given above and indicate that the films are not strained. It is worth noting that, to the best of our knowledge, epitaxial growth of AlFeO$_3$ films had not been reported earlier. Figure 1(b) shows a Q-plot displaying six {013} reflections of the ε-Fe$_2$O$_3$/AlFeO$_3$ film with three equivalent {110} STO reflections providing the in-plane orientation of the substrate for reference. This indicates the epitaxial growth of ε-Fe$_2$O$_3$ in six domains rotated 60° to each other as a result of the six-fold symmetry of the STO(111) surface (see Supporting Information for further details). The in-plane orientation of one of the domains is [100] ε-Fe$_2$O$_3$ ∥ [1-21] STO as displayed in the VESTA-constructed sketches[29] of Figure 1(c) and 1(d). In the case of AlFeO$_3$ films we obtained an equivalent pole figure which indicates that the buffer layer presents the same type of epitaxial growth, pointing out that ε-Fe$_2$O$_3$ grows cube-on-cube on AlFeO$_3$. Interestingly, the mismatch of an ε-Fe$_2$O$_3$ film directly grown on STO(111) is about -6 % in both directions whereas for AlFeO$_3$ decreases to about - 4% along [100] and -3 % along [010], indicating that on STO (111) the growth of AlFeO$_3$ is more favorable the growth of ε-Fe$_2$O$_3$. This observation fully justifies the selection of AlFeO$_3$ to be grown on STO(111) and its use as buffer layer for subsequent ε-Fe$_2$O$_3$ growth.

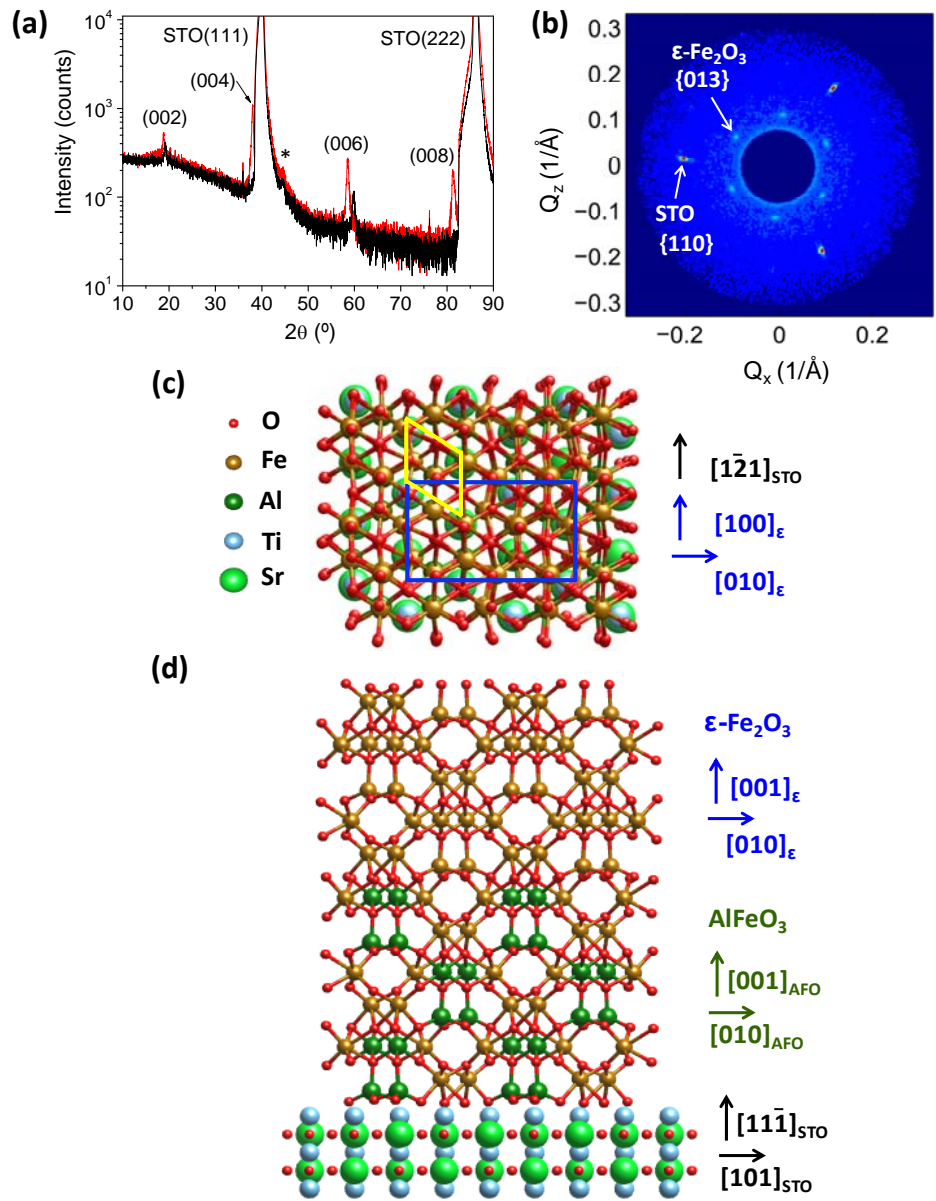

**Figure 1.** **(a)** XRD θ–2θ patterns. The black pattern corresponds to an AlFeO$_3$//Nb:STO(111) film with a thickness of 34 nm and the red pattern is from a ε-Fe$_2$O$_3$/AlFeO$_3$//Nb:STO(111) heterostructure with a 54 nm thick Fe$_2$O$_3$ film on a 8 nm thick AlFeO$_3$ buffer layer. The low intensity feature marked by an asterisk is of instrumental origin.[28] **(b)** Q-plot displaying the STO {110} and ε-Fe$_2$O$_3$ {013} reflections. **(c)** Schematic top view of the ε- Fe$_2$O$_3$/AlFeO$_3$//Nb:STO(111) heterostructure displaying the projections of the cell parameters of STO (in yellow) and ε-Fe$_2$O$_3$ (in blue) on the STO(111) surface, according to the epitaxial relationships deduced from the pole figure of panel (b) . **(d)** Schematic cross-sectional view of the ε-Fe$_2$O$_3$/AlFeO$_3$//Nb:STO(111) heterostructure with the epitaxial relationships deduced from the pole figure of panel (b).

Further insights into the microstructure of the ε-Fe$_2$O$_3$/AlFeO$_3$//Nb:STO(111) heterostructure were obtained by transmission electron microscopy (TEM). Figure 2(a) shows a cross-section bright-field TEM image of the ε-Fe$_2$O$_3$ layer, indicating that the film is continuous over the lateral length and showing the presence of three-dimensional islands with flat top surface and lateral size around 50 nm. Tilting the sample off-axis unveils the presence the AlFeO$_3$ layer grown onto the STO substrate with a thickness of about 8 nm, marked by a yellow arrow in the lower panel of Figure 2(a). The diffraction spots of the corresponding selected-area electron-diffraction (SAED) pattern can be indexed by considering the superposition of the diffraction patterns obtained along the zone axes [010] and [011] of ε-Fe$_2$O$_3$ and STO respectively, confirming the epitaxial growth of the ε-Fe$_2$O$_3$ film on STO as well as their epitaxial relationship. From the (200) and (004) diffraction spots of ε-Fe$_2$O$_3$, labelled in yellow in Figure 2(b), we have obtained the in-plane and out-of-plane cell parameters $a_\varepsilon$=5.05(5) Å and $c_\varepsilon$=9.47(8) Å, respectively. These values are in good agreement with the values reported for ε-Fe$_2$O$_3$ nanoparticles[30] and the out of plane cell parameter measured by XRD. The SAED pattern (Figure 2(b)) shows a splitting of the film and the substrate diffraction spots. This is clearly visible for the ε-Fe$_2$O$_3$ (-203) and STO (002) reflections marked by a yellow arrow and presented in an enlarged image, which confirm the relaxation of the ε-Fe$_2$O$_3$ film on the STO(111) substrate. The high resolution cross-sectional TEM image of Figure 2(c) shows a continuous and sharp ε-Fe$_2$O$_3$/AlFeO$_3$ interface further illustrating the quality of the epitaxial growth. In Figure 2(c), regions of distinctive contrast can be observed; these are fingerprints of the existence of domains with different in-plane orientation as described above. Topographic AFM images of the film surfaces (see Supporting Information, Figure S1) reveal a cluster-like morphology in agreement with the TEM observations. A moderately flat surface coexists with three-dimensional islands 10-35 nm in height.

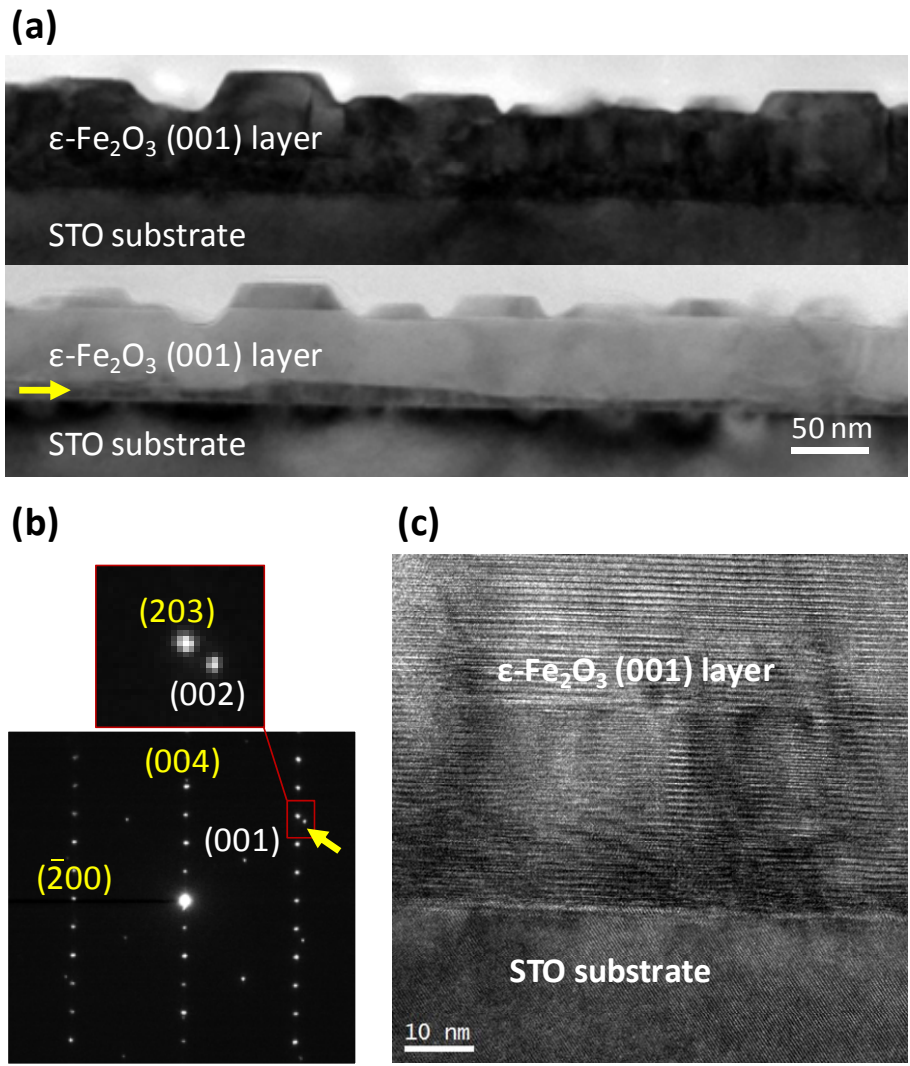

**Figure 2.** (a) Two cross-section bright-field TEM images of the ε-Fe$_2$O$_3$/AlFeO$_3$//STO(111) sample, one viewed along the [011] zone axis of STO (upper panel), and the other slightly tilted in order to disclose the presence of the AlFeO$_3$ layer, which appears as the dark stripe indicated with the yellow arrow. (b) SAED pattern along the [010] ε-Fe$_2$O$_3$ and [011] STO zone axes. The red square marks a region of the SAED pattern enlarged in the upper inset. The indexed diffraction spots of the ε-Fe$_2$O$_3$ layer and the substrate appear in yellow and white, respectively. The yellow arrow indicates the ε-Fe$_2$O$_3$ (203) and STO (002) reflections, illustrating that the film is relaxed both in-plane and out of plane. (c) High-resolution cross-section TEM image of the ε-Fe$_2$O$_3$/AlFeO$_3$//STO(111) heterostructure.

The in-plane magnetic field dependence of the film magnetization at room temperature shows a hysteresis loop typical of a ferromagnetic material (Figure 3(a)). The diamagnetic contribution from the substrate and the contribution of AlFeO$_3$ the buffer layer, which is paramagnetic at room temperature,[25] have been subtracted from the measured magnetic moment. It turns out that the ε-Fe$_2$O$_3$ film has a magnetic remanence and saturation magnetization of about 40 emu/cm$^3$ and 100 emu/cm$^3$, respectively, which are in good agreement with previous magnetic studies on ε-Fe$_2$O$_3$ nanoparticles.[16b, 30] The coercive field ~3 kOe is significantly reduced compared to previously reported values.[16b, 30] The step in the magnetization data observed at low field was also found in earlier magnetic measurements of pure ε-Fe$_2$O$_3$ [16b, 30] nanoparticles and related materials.[10b] These agreements, together with the measured cell parameters indicate that the structure and cation ordering of ε-Fe$_2$O$_3$ thin films mimic those of the nanoparticles. Stepped magnetization curves have been found in other ferrimagnetic oxides. This is the case of spinel ferrites where similar loops had been long ago reported in polycrystalline and single crystalline samples,[31] and more recently even in epitaxial thin films.[32] Although initially interpreted as anisotropy effects due to field annealing[31] the steps in the magnetization loops can have other microstructural causes, such as antiphase boundaries.[32] Even though the magnetization steps are a quite common observation in spinel ferrites, these phenomena are still not fully understood. The fact that both the spinel phase and ε-Fe$_2$O$_3$ are ferrimagnetic oxides with different magnetic sublattices displaying competing magnetic interactions suggests that a similar mechanism accounting for the stepped magnetization loop may hold in both systems.

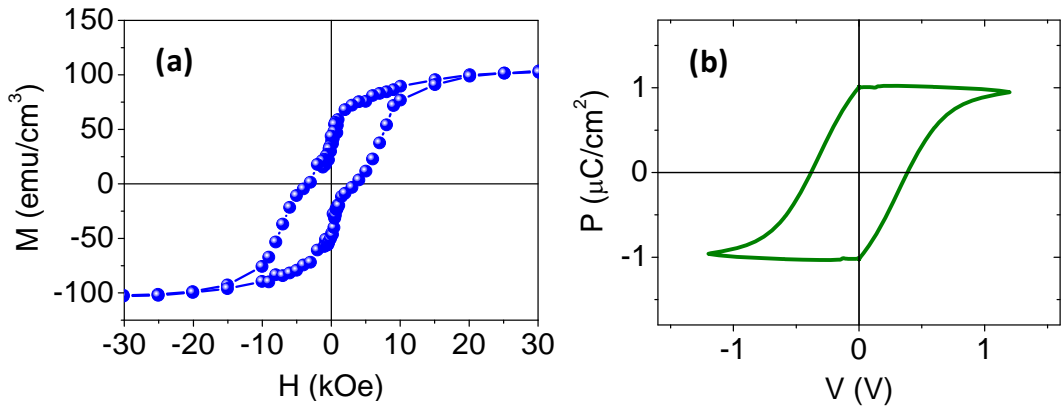

**Figure 3. (a)** Magnetization hysteresis loop at 300 K of ε-Fe$_2$O$_3$/AlFeO$_3$//Nb:STO(111) measured with the magnetic field applied in-plane. The data has been corrected for the diamagnetic and paramagnetic contributions from the substrate and the AlFeO$_3$ buffer layer, respectively. **(b)** Ferroelectric hysteresis loop measured at 300 K and 10 Hz with 100 ms of delay time.

We turn now to the dielectric properties of the films. The room-temperature polarization was determined by performing macroscopic ferroelectric measurements. The dependence of the film polarization on the applied voltage (Figure 3(b)) was obtained from the time integration of the purely ferroelectric displacement currents extracted from the application of the positive-up-negative-down (PUND) sequence of voltage pulses at different frequencies (see Supporting Information, Figure S2). The figure displays a characteristic fully saturated ferroelectric loop with a remanent polarization of about ~ 1 μC/cm$^2$ and a coercive voltage ~ 0.4 V, which corresponds to a coercive field of about 60 kV/cm. We notice that, due to the (00*l*) texture of the film, the polarization measurements have been performed along the ε-Fe$_2$O$_3$ polar axis (c-axis in the Pna2$_1$ space group setting).

The ferroelectric and piezoelectric responses of the films were also studied by piezo-force microscopy. Typical PFM hysteresis loops, i.e. amplitude (proportional to the piezoelectric coefficient d$_{33}$) and phase Φ versus applied voltage (V) are shown in Figure 4. The observed 180° phase contrast upon cycling V and the butterfly loop in d$_{33}$

(V) reveal the ferroelectric switching. The coercive voltage obtained using macroscopic contacts (Figure 3(b)) is smaller than the value recorded using the PFM and seems to be related to the different geometries (parallel *vs* point contact) and electric field distributions. It can be also observed the presence of sizeable imprint most probably due to the asymmetric work-function of the used electrodes (Nb:STO substrate and conductive tip).

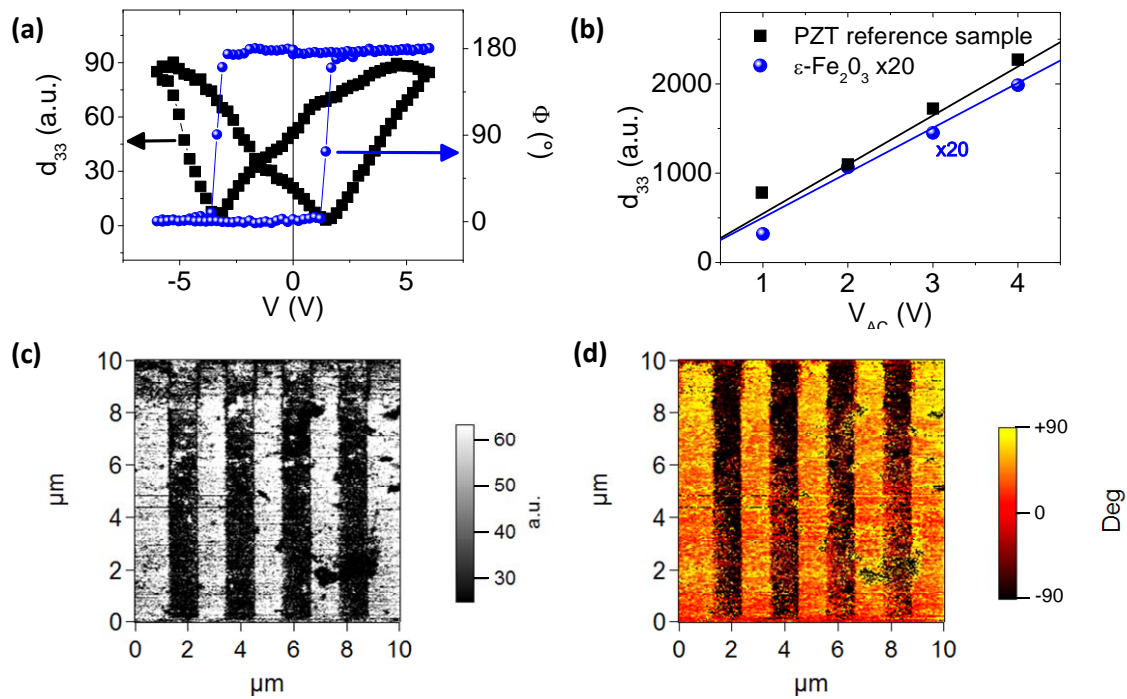

**Figure 4.** **(a)** Piezo-amplitude ((black symbols, left axis) and piezo-phase (blue symbols, right axis) loops versus applied voltage, measured with an AC driving voltage $V_{ac}$ = 0.5 V. **(b)** Piezo-signal response of the ε-Fe$_2$O$_3$/AlFeO$_3$ films as a function of $V_{ac}$ (blue circles), compared with a reference PZT sample with $d_{33}$ = 54 pm/V (black squares). The signal of the ε-Fe$_2$O$_3$/AlFeO$_3$ films has been multiplied by 20, therefore its piezoelectric coefficient can be estimated to be around 2.7 pm/V. **(c)** Piezo-signal and **(d)** piezo-phase images recorded at $V_{ac}$ = 0.5 V, after poling with +7 V (bright) and -7 V (dark).

The dependence of the PFM amplitude on excitation $V_{ac}$ is displayed in Figure 4(b) and shows a linear dependency as expected for a typical piezoelectric material. As mentioned below in the Experimental section, the the dual AC resonance tracking (DART) mode used to measure the PFM response precludes obtaining absolute values of the piezoelectric coefficients, but we can compare the measured piezoresponse to that of a sample comprising a reference ferroelectric layer. Thus, in Figure 4(b) we show the DART piezoresponse of a reference lead zirconium titanate (PZT) film with $d_{33}$ = 54 pm/V,[33] measured consecutively to the ε-Fe$_2$O$_3$/AlFeO$_3$ film using the same acquisition parameters. By comparing the response of both films we can estimate a $d_{33}$ coefficient of the ε-Fe$_2$O$_3$/AlFeO$_3$ film of about 2.7 pm/V. This value is similar to that reported for Ga$_{1-x}$Fe$_{1+x}$O$_3$ single crystals.[18a]

Amplitude and phase of the piezoresponse signal after poling the sample with +7 and -7 V are displayed in Figure 4(c,d), respectively. Both images show a sizeable contrast, with phase contrast of about 180°. The contrast in the piezo-amplitude response image occurs because of the presence of ferroelectric imprint which causes two remanent values for $d_{33}$, as can be seen in Figure 4(a). We have observed that the written domains are not affected by short-circuiting the sample surface and the bottom electrode through the cantilever (see Supporting Information, Figure S3) and persist even several weeks after being written (see Supporting Information, Figure S4).

The intrinsic ferroelectric nature of ε-Fe$_2$O$_3$ has been confirmed by performing additional PFM measurements on ε-Fe$_2$O$_3$ directly grown on a thin metallic SrRuO$_3$ bottom electrode in a ε-Fe$_2$O$_3$/SrRuO$_3$/AlFeO$_3$//STO(111) stack (see Supporting Information, Figure S5 and S6). In site of presenting an ε-Fe$_2$O$_3$ layer which is not as good as in the ε-Fe$_2$O$_3$/AlFeO$_3$//Nb:STO(111) heterostructures and a larger film leakage, the measurements on ε-Fe$_2$O$_3$/SrRuO$_3$/AlFeO$_3$//STO(111) show a response fully consistent with that of Figure 4 and provide an additional proof of the room-temperature ferroelectricity of the ε-Fe$_2$O$_3$ films.

Having confirmed the room-temperature ferroelectric character of ε-Fe$_2$O$_3$ thin films it could be useful to compare its remanent polarization (~ 1 μC/cm$^2$) with values reported for isostructural oxides. Sharma et al.[10b] and Thomasson et al.[12] recently measured room-temperature ferroelectric loops of GaFeO$_3$ and Ga$_{0.6}$Fe$_{1.4}$O$_3$:Mg films respectively, obtaining a saturation polarization of only ~ 0.1 μC/cm$^2$. Similarly, pyroelectric measurements on non-oriented Ga(Al)FeO$_3$ ceramics, at low temperature,[34] suggested a polarization not exceeding 0.3 μC/cm$^2$. We notice that all these reported values are significantly smaller than the ~ 2.5 μC/cm$^2$ first estimated by Arima[11] or the 25 -33 μC/cm$^2$ evaluated by D. Stoeffler and Mukherjee et al.[35] for these compounds.

D. Stoeffler showed that a simple point charge model can be used to estimate the polarization of GaFeO$_3$ if the displacements of the ions to the actual non-centrosymmetric structure from a centrosymmetric one are known.[35a] Following the same approach, we used the PSEUDO code[36] to find the centrosymmetric structures related to the available room-temperature structural refinements of ε-Fe$_2$O$_3$ in the non-centrosymmetric Pna2$_1$ space group.[27, 30, 37] In agreement with Stoeffler, the "nearest" centrosymmetric super group is found to be the Pnna and the atomic positions of the actual ε-Fe$_2$O$_3$ structures are compared to those of the atoms in the Pna2$_1$ group and used to calculate the polarization using a point charge model. We obtained spontaneous polarizations ranging from 5.4 to 23.3 μC/cm$^2$, depending on the precise values of atomic positions of the ε-Fe$_2$O$_3$ structure.[27, 30, 37] Our experimental value satisfactorily approaches the lower bound of these theoretical estimates. However, this comparison also shows the extreme sensitivity of polarization to the real atomic positions.

Finally, magnetolectric coupling has been explored by means of magnetocapacitance measurements. In Figure 5(a), the magnetocapacitance response [MC = (C(H)-C(0))/C(0)], where C is the room-temperature capacitance of the films stack, is displayed at two frequencies. As shown in Figure 5(b), the magnetolosses [ML = (tanδ(H)-tanδ(0))/tanδ(0)] at these frequencies do not show any significant variation with magnetic field, thus suggesting negligible extrinsic contributions to the magnetocapcitance and thus MC(H) can be taken as an estimate of the change of the film permittivity with magnetic field. The MC(H) data in Figure 5(a) display a clear hysteresis that closely follows the magnetization loops shown in Figure 3. In fact, the minima of MC(H), occurring below ±4 kOe are comparable with the coercive fields observed in the magnetization loops. This agreement indicates that there is a coupling between the electric and magnetic orders. However, the absolute variation of the capacitance, around $10^{-6}$ %/Oe, is rather low. The corresponding magnetoelectric coupling $\chi_E$, evaluated using the relation MC(H) = $(\chi_E/E_0)$H,[38] where $E_0$ is the applied excitation electric field is 0.2 mV/cm·Oe, around three orders of magnitude lower than values recently reported for some hexaferrites (assuming constant permittivity).[39] Although the present measurements do not allow determining the microscopic nature of the observed magnetoelectric coupling, the observed small $\chi_E$ value is primarily dictated by the hard magnetic character of the ε-Fe$_2$O$_3$, thus implying that larger magnetic fields (around 5 kOe) are required to significantly modify the magnetic texture of this iron oxide.

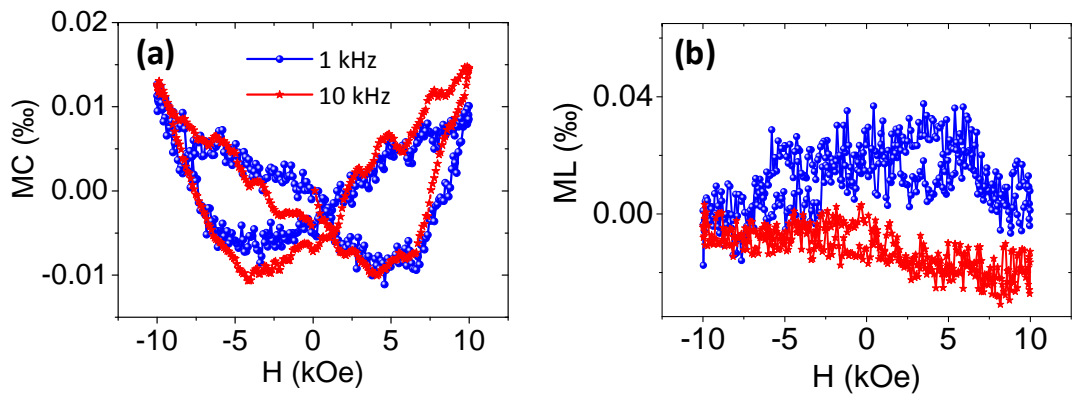

**Figure 5.** (a) Magnetocapacitance (MC = (C(H)-C(0))/C(0)) versus applied magnetic field at 1 and 10 kHz and $V_{ac}$ = 100 mV. (b) Magnetolosses (ML = (tanδ(H)-tanδ(0))/tanδ(0)) versus applied magnetic field at 1 and 10 kHz and $V_{ac}$ = 100 mV.

**Conclusions**

In summary, we have shown that the metastable ε-Fe$_2$O$_3$ oxide can be epitaxially grown on perovskite substrates by using isostructural oxides (AlFeO$_3$, in the present work) as seed layer, thus allowing a robust and high-yield growth of ε-Fe$_2$O$_3$ epitaxial films. Room-temperature piezoforce microscopy and macroscopic polarization measurements have revealed that ε-Fe$_2$O$_3$ films display a typical ferroelectric switching with significant polarization (~1 μC/cm$^2$) and low switching voltages. This, along with its well know ferrimagnetic character, demonstrates that ε-Fe$_2$O$_3$ is a new and unique intrinsic room temperature multiferroic material, which could find novel applications. For instance, we can expect that optical magnetoelectric effects, analogous to those reported in GaFeO$_3$,[40] will occur at room-temperature in ε-Fe$_2$O$_3$ films and can be voltage-controlled via polarization switching. In a scenario of multiferroic materials integration into real devices, the chemical simplicity of ε-Fe$_2$O$_3$ represents a major advantage to overcome the challenges of stoichiometry and phase purity control that might be cumbersome in all other known room-temperature multiferroics.

**Experimental Section**

*Preparation of the films*: AlFeO$_3$ and ε-Fe$_2$O$_3$ films were grown by pulsed laser deposition (PLD) using a KrF excimer laser (248 nm wavelength, 25 ns pulse duration) at 5 Hz repetition rate on Nb-doped (0.5%) (111)STO single crystals. The laser beam was focused at about 1.5 J/cm$^2$ on Fe$_2$O$_3$ and AlFeO$_3$ ceramic targets prepared by sintering α-Al$_2$O$_3$ and α-Fe$_2$O$_3$ powders. The substrates, placed at a distance around 5 cm from the targets, were heated to high temperature under a dynamic oxygen pressure of 10 Pa. Several tens of AlFeO$_3$ films and ε-Fe$_2$O$_3$/AlFeO$_3$ heterostructures were grown. The preparation of the heterostructures could be achieved in the wide 750 °C-900 °C temperature range and turned out to be fully reproducible, indicating the robustness of our strategy to stabilize ε-Fe$_2$O$_3$ films. Further details of the relationships between growth conditions and structure will be reported elsewhere.[41] The heterostructure selected for the dielectric and magnetic characterization and extensively described in the manuscript consisted of a ε-Fe$_2$O$_3$ film with a thickness around 54 nm grown on an 8 nm thick buffer layer of AlFeO$_3$, both deposited at 825 °C (denoted ε-Fe$_2$O$_3$/AlFeO$_3$). Structural data for a 34 nm thick single AlFeO$_3$ layer on Nb:STO(111), is also reported.

*Structural characterization*: The structure and microstructure of the films were studied by XRD using Cu- Kα radiation in a Rigaku, a Panalytical MRD X'Pert Pro and a Bruker-AXS D8 Advance diffractometers, transmission electron microscopy in a FEI Tecnai G$^2$ F20 operated at 200 kV and atomic force microscopy (AFM) using an Agilent 5100 SPM apparatus. The magnetic characterization was performed with a Quantum Design SQUID magnetometer. Raw measurements of the magnetization *Vs* applied magnetic field were corrected for the paramagnetic contributions from the substrate and AlFeO$_3$ layer by substracting the linear response measured at high fields. For the dielectric characterization of the films, Pt electrodes (~60 nm thickness, and ~0.17 mm$^2$ area) were sputtered *ex-situ* on the film surface through a suitable mask.

*Characterization of film properties*: PFM measurements were performed with a MFP-3D Asylum Research microscope. MikroMasch® silicon cantilevers with TiPt coating (NSC35), and AppNano Co. with Pt coating (ANSCM-PT) were used with no difference in the obtained results. To achieve better sensitivity, the DART method was employed.[42] PFM voltage hysteresis loops were always performed at remanence, using a dwell time of 100 ms. The quantification of the piezo coefficient using DART is difficult due to the simultaneous variation of measurement frequency and the variation of the maxima of the resonance amplitude while measuring; consequently arbitrary units (a.u.) are indicated in the amplitude of the piezoresponse. Still, the piezoelectric coefficient of the ε-Fe$_2$O$_3$/AlFeO$_3$ heterostructure was estimated by comparison with a PZT calibration sample measured in the same set-up and experimental conditions.

Macroscopic Ferroelectric/Dielectric measurements were performed using a planar capacitor top-to-top electrode configuration.[43] Capacitance was measured using a HP4182 LF impedance analyzer (Agilent Co.). The polarization was evaluated by measuring the dynamic P-V hysteresis loops using an AixAcct TF analyzer 2000 (aixACCT Systems GmbH) employing the PUND technique,[44] as described in a previous work.[45] Magnetic field (applied in-plane) in magnetocapacitance measurements have been controlled inserting the samples in a Physical Property Measurement System from Quantum Design using a special designed inset.

**Acknowledgements**
This research was partially funded by the Spanish Government (MAT2012-35324, MAT2011-29269-CO3, CONSOLIDER-Nanoselect-CSD2007-00041, the RyC-2012-11709 contract of J. G. and RyC-2009-04335 contract of M. G.), the Generalitat de Catalunya (2009SGR203, 2009SGR00376 and the Beatriu de Pinós 2011BP-A00220 postdoctoral grant of I.F.) and the European Commission (FP7-Marie Curie Actions, PCIG09-GA-2011–294168 grant of M. G.) as well as the COST Action MP1202. Authors acknowledge B. Ballesteros, J. Santiso and P. García from Electron Microscopy and Diffraction Divisions of the Catalan Institute of Nanoscience and Nanotechnology (ICN2) for offering access to their instruments and expertise. Xavi Martí is deeply acknowledged for preparing Figure 1b of this work.

Supporting Information

**Multiferroic Iron Oxide Thin Films at Room-Temperature**

*Martí Gich\*, Ignasi Fina, Alessio Morelli, Florencio Sánchez, Marin Alexe, Jaume Gàzquez, Josep Fontcuberta and Anna Roig*

# Epitaxial relationships

The diffraction patterns of ε-$Fe_2O_3$/AlFeO$_3$ films are largely dominated by the ε-$Fe_2O_3$ contribution owing to the very small diffracted intensity of the AlFeO$_3$ buffer (8 nm). In Figure 1(b) of the manuscript, as expected from the (00$l$) out-of-plane texture of the film and the (111) orientation of the STO substrate, the Q plot displays the {013} ε-$Fe_2O_3$ and {110} STO reflections were recorded at tilt angles slightly below 20° and at about 35°, respectively. Since the {013} ε-$Fe_2O_3$ family of planes only consists of the (013) and (01-3) planes, whose reflections have to appear 180° apart in the reciprocal space, the presence of six {013} ε-$Fe_2O_3$ reflections implies the existence of three or six in-plane domains which are compatible with six-fold symmetry of the STO(111) surface. The epitaxial relationships between film and substrate can be worked out from the observation that in Fig. 1(b) the {110} STO reflections are 30° apart from the closest {013} ε-$Fe_2O_3$ reflections. The projections of [013] ε-$Fe_2O_3$ and [110] STO on the (001) ε-$Fe_2O_3$ and (111) STO planes are parallel to [010] ε-$Fe_2O_3$ and [11-2] STO directions, respectively. Since the angle between the [11-2] STO and [10-1] STO directions on (111) STO is also 30°, it can be concluded that the in-plane orientation of one of the domains is [010] ε-$Fe_2O_3$ ∥ [10-1] STO (*i.e.* with [100] ε-$Fe_2O_3$ ∥ [1-21] STO). In the case of AlFeO$_3$ films we obtained an equivalent Q-plot diagram which indicates that the buffer layer presents the same type of epitaxial growth, pointing out that ε-$Fe_2O_3$ grows cube-on-cube on AlFeO$_3$. Figure 1(c) shows the projections of the STO and ε-$Fe_2O_3$ unit cells on the STO(111) plane from which one can calculate the lattice mismatch between the film and the substrate along the [100] and [010] film directions according to ($\sqrt{6}a_{STO}$–2$a_{film}$)/2$a_{film}$ and (3$\sqrt{2}a_{STO}$–2$b_{film}$)/2$b_{film}$. From these, the mismatch of an ε-$Fe_2O_3$ film directly grown on STO(111) is about -6 % in both directions whereas for AlFeO$_3$ decreases to about -4% along [100] and -3 % along [010], indicating that on STO (111) the growth of AlFeO$_3$ is more favorable than the growth of ε-$Fe_2O_3$.

## AFM topography

Figure S1(a) displays the topographic AFM image of the very same ε-$Fe_2O_3$(54 nm)/$AlFeO_3$(8 nm)//Nb:$SrTiO_3$(111) sample reported in the main manuscript and the height profile along the marked line (Figure S1(b)). The film shows a rather flat surface, but there is a random distribution of high three-dimensional islands with heights up to 35 nm. The overall rms roughness is about 3 nm.

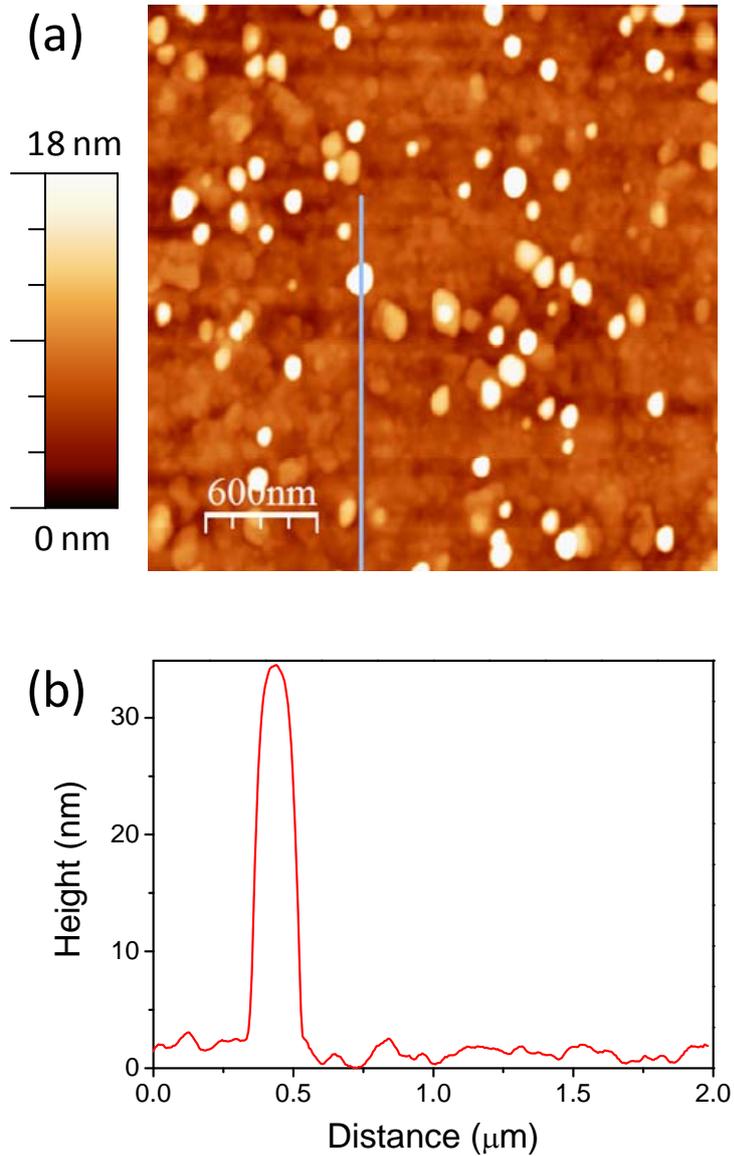

**Figure S1. (a)** Topographic AFM image and **(b)** height profile along the marked line in the topographic image.

**Ferroelectric switching of macroscopic capacitors**

Figure S2(a) shows the current versus voltage (I-V) characteristics recorded after the application of each voltage pulse from the PUND sequence, shown schematically in the inset, except for the first pulse, which switches the sample into a negative initial state. The switching current curves for the first pulses of positive and negative polarities (P and N) comprises both ferroelectric (*i.e.* displacement current arising from the switching of ferroelectric domains) and non-ferroelectric contributions (*i.e.* displacement current due to the charging of the material and leakage current) whereas for the subsequent pulses with the same polarity (U and D) only the non-ferroelectric contributions are present in the I-V curves.

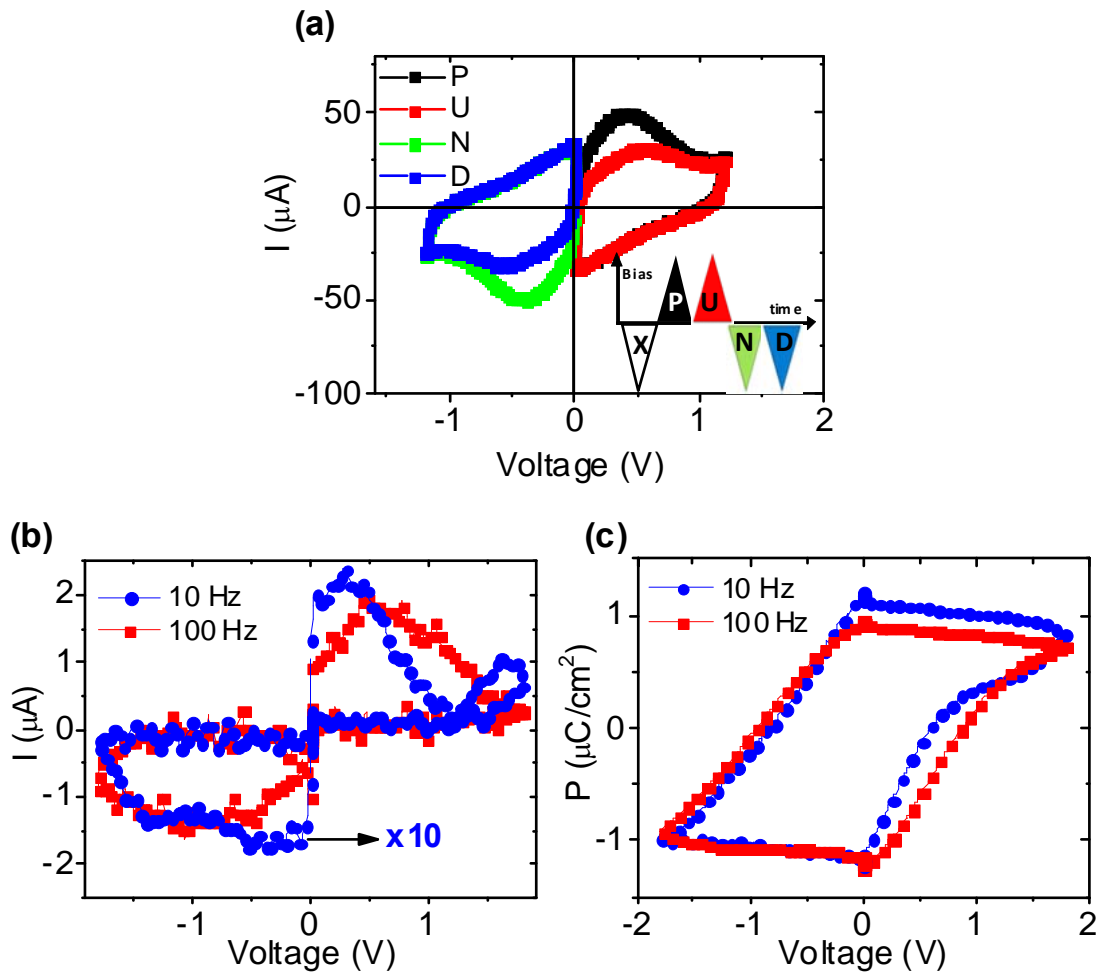

**Figure S2**. (a) Ferroelectric measurements of the ε-$Fe_2O_3$(54 nm)/$AlFeO_3$(8 nm)//Nb:$SrTiO_3$(111) heterostructure, i.e. current–voltage charateristics, obtained using PUND technique at 300 K and 10 Hz. The inset schematically represents the voltage bias sequence of the PUND method. (b) Displacement current versus voltage response at 300 K measured at 10 and 100 Hz. (c) Ferroelectric loops obtained from the current-voltage measurements in panel (b).

The P and N curves of Figure S2(a) clearly display ferroelectric switching current peaks. Note the significant broadness of the current peaks, resulting from a distribution of coercive fields across the film which in the present case is most probably due to the broadness of grain sizes observed in the film surface characterization (see Figure S1). Figure 3(b) of the manuscript depicts the time integration of the purely ferroelectric displacement currents obtained from the differences P-U and N-D from the data of Figure S2(a).

Figure S2(b) shows that the displacement currents recorded in the ε-Fe$_2$O$_3$(54 nm)/AlFeO$_3$(8 nm)//Nb:SrTiO3(111) heterostructure increase linearly with the measurement frequency. Figure S2(c) shows an overlap of the polarization versus voltage loops obtained from the time-integration of these signals measured at 10 and 100 Hz, as expected from the response of the polarization to an applied electricfield in a ferroelectric material.

**PFM switching**

In order to provide an additional confirmation of the ferroelectric character of the ε-Fe$_2$O$_3$/AlFeO$_3$//Nb:STO(111) sample, we have performed the following experiment. First, the sample was poled by using ± 7 V and the PFM image was recorded using V$_{ac}$ = 2 V. In Figure S3(a) and S3(b) we show the amplitude and phase images, respectively, where the written domains can be seen as vertical stripes. In Figure S3(a) these appear as white/black regions. Next, the AFM tip and the bottom electrode were short-circuited and the sample surface was scanned again at V$_{ac}$ = 0. These images, superimposed in Figure S3 bounded by yellow dashed lines, have no contrast as expected by the screening of the film surface charges and thus the electrostatic force on the cantilever does not show contrast on top of up/down poled regions. Next, the tip-bottom electrode short-circuit is removed and the piezoresponse is measured again using V$_{ac}$ = 2 V. The images obtained, superimposed in Figure S3 bounded by white dashed lines, show the same domain structure as initially written, thus confirming that the written domains have a surface charge related to film polarization. Note that the use of large excitation contrast (V$_{ac}$ = 2 V) lowers the contrast in the phase-PFM images due to UP and DOWN domains signal convolution.

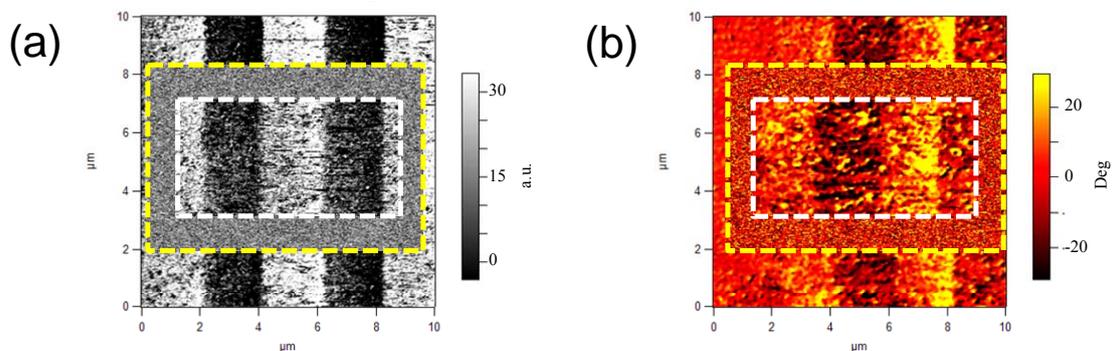

**Figure S**3. **(a, b)** Piezo-amplitude, and piezo-phase images. The stripes were written by using ± 7 V and the piezoresponse recorded using using V$_{ac}$ = 2 V. Amplitude and phase (Figure S3(a) and Figure S3(b)) were recorded inside the regions framed by yellow

dashed lines while short-circuiting the bottom electrode and the cantilever and superimposed in the figure. Amplitude and phase were later recorded inside the rectangular region framed by white dashed lines after removing the tip-bottom electrode shortcircuit, with $V_{ac} = 2$ V.

Figure S4 shows the piezo-response (amplitude) image of domains, three weeks after being written with -7V and +7V.

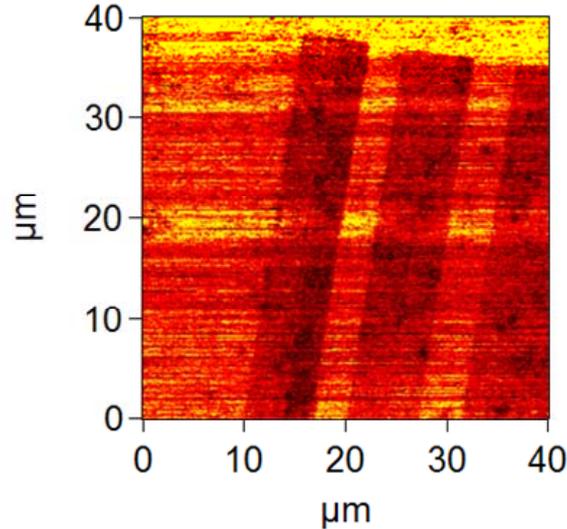

**Figure S4.** Piezo-signal image recorded 3 weeks after being written.

PFM characterization was also performed in a ε-Fe$_2$O$_3$(50 nm)/SrRuO$_3$(100 nm)/AlFeO$_3$ (8 nm) heterostructure deposited on an undoped SrTiO$_3$(111) substrate. To perform PFM measurements of the ε-Fe$_2$O$_3$ layer, the sample was contacted by grounding the SrRuO$_3$ electrode. Figure S5(a,b) show the hysteretic piezo-amplitude signal (a) and the 180° piezo-phase (b) obtained using $V_{ac}$=0.5 V. Both sets of data show clear ferroelectric response. In fact, except for the larger imprint observed here, the results of Figure S5 are similar to those recorded for the ε-Fe$_2$O$_3$/AlFeO$_3$//Nb:STO(111) sample shown in the manuscript.

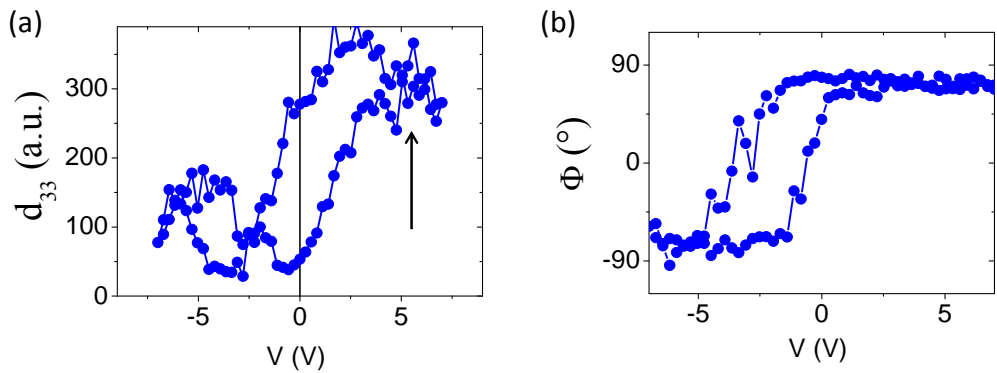

**Figure S5** PFM characterization of an ε-Fe$_2$O$_3$(50 nm)/SrRuO$_3$(100 nm)/AlFeO$_3$(8 nm) heterostructure deposited on an undoped SrTiO$_3$(111) **(a, b)** Piezo-signal and piezo-phase loops versus electric field with $V_{ac}$ = 0.5V.

Reliable macroscopic electric measurements on this sample were not possible due to the presence of an important leakage, as revealed by the saturation of the piezo-amplitude signal at high voltages indicated with an arrow in Figure S5(a). Most probably, this is because this film shows a pronounced granular morphology with height differences of tens of nanometers, as it can be seen in the AFM topography image and profile Figure S6(a) and S6(b), which favours short-circuits at the grain boundaries.

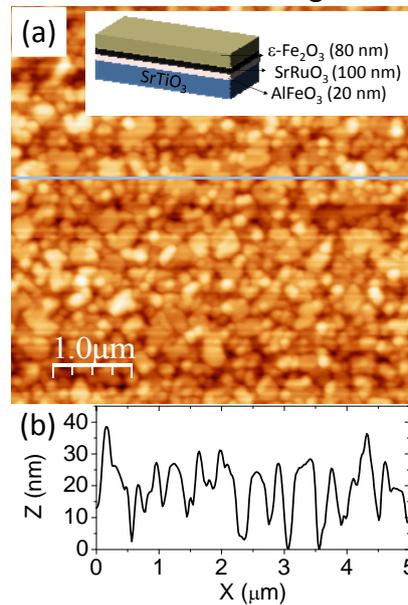

**Figure S6. (a)** AFM topographic image along the marked line. **(b)** Height profile of the ε-$Fe_2O_3$/$SrRuO_3$/$AlFeO_3$//STO(111) sample.